\documentclass{elsart}

\usepackage{amssymb}

\newcommand{\bec}[1]{\mbox{\boldmath $ #1$}}
\usepackage{graphicx}

\begin{document}

\begin{frontmatter}



\title{New mechanism of generation of large-scale magnetic field
in a sheared turbulent plasma}

\author{N. Kleeorin and I. Rogachevskii}

\address{Department of Mechanical Engineering,
 Ben-Gurion University of Negev, POB 653, 84105 Beer-Sheva, Israel}

\begin{abstract}
A review of recent studies on a new mechanism of generation of
large-scale magnetic field in a sheared turbulent plasma is
presented. This mechanism is associated with the shear-current
effect which is related to the ${\bf W} {\bf \times} {\bf J}$-term
in the mean electromotive force. This effect causes the generation
of the large-scale magnetic field even in a nonrotating and
nonhelical homogeneous sheared turbulent convection whereby the
$\alpha$ effect vanishes (where ${\bf W}$ is the mean vorticity due
to the large-scale shear motions and ${\bf J}$ is the mean electric
current). It is found that turbulent convection promotes the
shear-current dynamo instability, i.e., the heat flux causes
positive contribution to the shear-current effect. However, there is
no dynamo action due to the shear-current effect for small
hydrodynamic and magnetic Reynolds numbers even in a turbulent
convection, if the spatial scaling for the turbulent correlation
time is $\tau(k) \propto k^{-2}$, where $k$ is the small-scale wave
number. We discuss here also the nonlinear mean-field dynamo due to
the shear-current effect and take into account the transport of
magnetic helicity as a dynamical nonlinearity. The magnetic helicity
flux strongly affects the magnetic field dynamics in the nonlinear
stage of the dynamo action. When the magnetic helicity flux is not
small, the saturated level of the mean magnetic field is of the
order of the equipartition field determined by the turbulent kinetic
energy. The obtained results are important for elucidation of origin
of the large-scale magnetic fields in astrophysical and cosmic
sheared turbulent plasma.
\end{abstract}


\end{frontmatter}

\section{Introduction}
\label{}

Turbulence with a large-scale velocity shear is a universal feature
in astrophysical plasmas. It has been recently recognized that in a
sheared turbulent plasma with high hydrodynamic and magnetic
Reynolds numbers a mean-field dynamo is possible even in a
nonhelical and nonrotating homogeneous turbulence whereby a kinetic
helicity and $\alpha$ effect vanish (see Rogachevskii and Kleeorin
2003; 2004; 2007; Brandenburg 2005; Brandenburg and Subramanian 2005c;
Rogachevskii et al. 2006a; 2006b). The large-scale velocity shear
produces anisotropy of turbulence with a nonzero background mean
vorticity ${\bf W} = \bec{\nabla} {\bf \times} {\bf U}$, where ${\bf
U}$ is the mean velocity. The dynamo instability in a sheared
turbulent plasma is related to the ${\bf W} {\bf \times} {\bf
J}$-term in the mean electromotive force, and it can be written in
the form $\bec{\cal E}^\delta \propto - l_0^2 \, {\bf W} {\bf
\times} (\bec{\nabla} {\bf \times} {\bf B}) \propto l_0^2 \, ({\bf
W} {\bf \cdot} {\bf \Lambda}^B) {\bf B}$, where $l_0$ is the maximum
scale of turbulent motions (the integral turbulent scale) and $ {\bf
\Lambda}^B = \bec{\nabla} {\bf B}^2 / 2{\bf B}^2 $ determines the
inhomogeneity of the mean original magnetic field ${\bf B}$. In a
sheared turbulent plasma the deformations of the original magnetic
field lines are caused by the upward and downward turbulent eddies,
and the inhomogeneity of the original mean magnetic field in the
shear-current dynamo breaks a symmetry between the influence of
upward and downward turbulent eddies on the mean magnetic field.
This creates the mean electric current ${\bf J}$ along the mean
magnetic field and produces the mean-field dynamo due to the
shear-current effect.

The goal of this communication is to review recent studies on the
new mechanism of generation of large-scale magnetic field due to the
shear-current effect in a sheared turbulent plasma. The mean-field
dynamo instability is saturated by the nonlinear effects. There are
two types of the nonlinear effects caused by algebraic and dynamic
nonlinearities. The effects of the mean magnetic field on the motion
of fluid and on the cross-helicity result in quenching of the mean
electromotive force which determines the algebraic nonlinearity. The
dynamical nonlinearity in the mean-field dynamo is caused by the
evolution of small-scale magnetic helicity, and it is of a great
importance due to the conservation law for the magnetic helicity in
turbulent plasma with very large magnetic Reynolds numbers (see,
e.g., Kleeorin and Rogachevskii 1999; Brandenburg and Subramanian
2005a; Rogachevskii et al. 2006b, and references therein). The
combined effect of the dynamic and algebraic nonlinearities
saturates the growth of the mean magnetic field.

The shear-current effect has been studied by Rogachevskii and
Kleeorin (2003; 2004; 2007) for large hydrodynamic and magnetic Reynolds
numbers using two different approaches: the spectral $\tau$
approximation (the third-order closure procedure) and the stochastic
calculus (the path integral approach in a turbulence with a finite
correlation time). A justification of the $\tau$ approximation for
different situations has been performed in numerical simulations and
analytical studies by Blackman and Field (2002); Field and Blackman
(2002); Brandenburg at al. (2004); Brandenburg and Subramanian
(2005a, 2005b); Sur et al. (2007).

\section{The shear-current effect}
\label{}

Let us consider a nonhelical and nonrotating homogeneous turbulent
plasma with a weak mean velocity shear, ${\bf U} = (0, Sx, 0)$ and
mean vorticity ${\bf W}= (0,0,S)$, where $(S \, \tau_0)^2 \ll 1$,
$\, \tau_0 = l_0 / \sqrt{\langle {\bf u}^2 \rangle}$ and ${\bf u}$
are the velocity fluctuations. The mean magnetic field ${\bf B}(t,z)
= (B_x, B_y, 0)$ in the kinematic approximation is determined by the
following equations
\begin{eqnarray}
{\partial B_x(t,z) \over \partial t} &=& - S \, l_0^2 \,
\sigma_{_{B}} \, B''_y + \eta_{_{T}} \, B''_x \;,
\label{E1}\\
{\partial B_y(t,z) \over \partial t} &=& S \, B_x + \eta_{_{T}} \,
B''_y \;,
\label{E2}
\end{eqnarray}
where $B''_i = \partial^2 B_i / \partial z^2 $, $\, \eta_{_{T}}$ is
the coefficient of turbulent magnetic diffusion and the
dimensionless parameter $\sigma_{_{B}}$ determines the shear-current
effect (Rogachevskii and Kleeorin 2003, 2004). In Eqs.~(\ref{E1})
and~(\ref{E2}) we have taken into account that $B_y \gg B_x$ since
$(S \, \tau_0)^2 \ll 1$. The first term $ \propto S B_x $ in
Eq.~(\ref{E2}) determines the stretching of the magnetic field $B_x$
by the shear motions, which produces the field $B_y$. The
interaction of the non-uniform magnetic field $B_y$ with the
background vorticity ${\bf W}$ produces the electric current along
the field $B_y$. This implies generation the field component $B_x$
due to the shear-current effect, which is determined by the first
term  $ \propto - \sigma_{_{B}} \, S \, l_0^2 \, B''_y $ in
Eq.~(\ref{E1}). This effect results in the large-scale dynamo
instability. The solution of Eqs.~(\ref{E1}) and~(\ref{E2}) we seek
for in the form $ \propto \exp(\gamma \, t + i K_z \, z)$, where the
growth rate $\gamma$ of the mean magnetic field due to the dynamo
instability is given by $\gamma = S \, l_0 \, \sqrt{\sigma_{_{B}}}
\, K_z - \eta_{_{T}} \, K_z^2 $. The necessary condition for the
dynamo instability is $\sigma_{_{B}} > 0$.

The shear-current dynamo instability depends on the spatial scaling
of the correlation time $\tau(k) \propto k^{-\mu}$ of the turbulent
velocity field, where $k$ is the small-scale wave number. In
particular, the shear-current dynamo in a non-convective turbulence
occurs when the exponent $\mu < 1$. For the Kolmogorov's type
turbulent convection, the exponent $\mu=2/3$ and $\sigma_{_{B}} = (4
/ 135) \, [1 + (6 / 7) \, a_\ast ]$, where the convective
contribution to the dynamo instability due to the shear-current
effect depends on the parameter $a_\ast = 2 \, g \, \tau_0 \, F^\ast
/ \langle {\bf u}^2 \rangle$. Here $F^\ast$ is an imposed vertical
heat flux which maintains the turbulent convection and ${\bf g}$  is
the acceleration of gravity. For a turbulent convection with a
scale-independent correlation time, the exponent $\mu=0$ and the
parameter $\sigma_{_{B}}$ is given by $\sigma_{_{B}} = (1 /15) \, [1
+ (9 / 7) \, a_\ast \, (1 + 3 \sin^2 \, \phi)]$, where $\phi$ is the
angle between the background mean vorticity ${\bf W}$ and ${\bf g}$.
Note that the turbulent convection promotes the shear-current dynamo
instability. In particular, the heat flux causes positive
contribution to the shear-current effect when $2 + 3 \, (2 - 3 \mu)
\, \sin^2 \, \phi >0$ (see Rogachevskii and Kleeorin, 2007).

However, for small hydrodynamic and magnetic Reynolds numbers, the
turbulent correlation time is of the order of $\tau(k) \propto
1/(\nu k^2)$ or $\tau(k) \propto 1/ (\eta k^2)$ depending on the
magnetic Prandtl number, i.e., $\tau(k) \propto k^{-2}$, where $\nu$
is the kinematic viscosity and $\eta$ is the magnetic diffusion due
to the electrical conductivity of the plasma. In this case $\mu=2$,
and the parameter $\sigma_{_{B}} < 0$ even in a turbulent
convection. This implies that for small hydrodynamic and magnetic
Reynolds numbers there is no dynamo action due to the shear-current
effect. This result is in agreement with the recent studies by
R\"{a}dler and Stepanov (2006) and R\"{u}diger and Kitchatinov
(2006), where the dynamo action  have not been found in non-helical
and non-rotating sheared non-convective turbulent plasma in the
framework of the second-order correlation approximation  (SOCA)  or
the first-order smoothing approximation (FOSA). This approximation
is valid only for small hydrodynamic Reynolds numbers. Even in a
highly conductivity limit (large magnetic Reynolds numbers), SOCA
can be valid only for small Strouhal numbers, while for large
hydrodynamic Reynolds numbers (for a developed turbulence), the
Strouhal number is 1.

Note that the standard approach (i.e., SOCA) cannot describe the
situation in principle. The reason is that the shear-current dynamo
requires a finite correlation time of turbulent velocity field, so
the delta-correlated version of SOCA fails. The application of the
path integral approach for the study of the shear-current dynamo
requires a finite correlation time of turbulent velocity field.
The shear-current dynamo is a phenomenon that results from the
interaction of the energy-containing-scale of turbulence with
large-scale shear, and the constraint is that the hydrodynamic and
magnetic Reynolds numbers should be not small at least. Therefore,
the SOCA-based approaches do not work properly to describe the
shear-current dynamo. Probably, the hydrodynamic and magnetic
Reynolds numbers can be of the order of unity and there is no need
for a developed inertial range in order to maintain the
shear-current dynamo.

\section{Nonlinear effects}
\label{}

In order to find the magnitude of the magnetic field, the nonlinear
effects must be taken into account. The nonlinear shear-current
dynamo have been studied by Rogachevskii and Kleeorin (2004);
Rogachevskii et al. (2006a; 2006b). The mean magnetic field is
determined by the following nonlinear equations
\begin{eqnarray}
&& {\partial B_x(t,z) \over \partial t} = - S \, l_0^2 \,
[\sigma_{_{B}}(B) \, B'_y]' - [\alpha_m(B) \, B_y]' + \eta_{_{T}} \,
B''_x \;,
\label{RM1} \\
&& {\partial B_y(t,z) \over \partial t} =  S \, B_x + \eta_{_{T}} \,
B''_y \;,
\label{RM2} \\
&& {\partial \chi_c(t,z) \over \partial t} - \kappa_{_{T}} \,
\chi_c'' + {\chi_{c} \over \tau_\chi} = - {1 \over 9 \, \pi \, \rho
\, \eta_{_{T}}} \, \bec{\cal E} \, {\bf \cdot} \, {\bf B} \;,
\label{RM3}
\end{eqnarray}
where $B'_i = \partial B_i / \partial z$, $\, \bec{\cal E} =
\alpha_m \, {\bf B} + S \, l_0^2 \, \sigma_{_{B}}(B) \, B'_y \, {\bf
e}_y - \eta_{_{T}} \, (\bec{\nabla} {\bf \times} {\bf B})$ is the
mean electromotive force, $\alpha_m = \chi_c(t,z) \, \Phi_{_{N}}(B)$
is the magnetic $\alpha$ effect, $\Phi_{_{N}}(B)$ is the quenching
function of the magnetic $\alpha$ effect and $\rho$ is the fluid
density. The function $\chi_{c}({\bf B})$ is related to the
small-scale current helicity $\langle {\bf b} {\bf \cdot}
(\bec{\nabla} {\bf \times} {\bf b}) \rangle$, where ${\bf b}$ are
the magnetic fluctuations. For a weakly inhomogeneous turbulent
plasma, the function $\chi_{c}$ is proportional to the small-scale
magnetic helicity. In Eq.~(\ref{RM3}) we use the simplest form of
the magnetic helicity flux, $\propto - \kappa_{_{T}} \bec{\nabla}
\chi_c$, where $\kappa_{_{T}}$ is the coefficient of the turbulent
diffusion of the magnetic helicity, $\tau_\chi = \tau_0 \, {\rm Rm}$
is the characteristic relaxation time of the small-scale magnetic
helicity and ${\rm Rm}$ is the magnetic Reynolds number.
Equations~(\ref{RM1}) and (\ref{RM2}) follow from the mean-field
induction equation, while Eq.~(\ref{RM3}) is derived using arguments
based on the magnetic helicity conservation law (see, e.g., Kleeorin
and Rogachevskii 1999; Brandenburg and Subramanian 2005a;
Rogachevskii et al. 2006b, and references therein). For large
magnetic Reynolds numbers the relaxation term $\chi_{c} / \tau_\chi$
in Eq.~(\ref{RM3}) can be neglected. For moderate values of the
magnetic Reynolds numbers this term has been taken into account by
Brandenburg and Subramanian (2005c); Rogachevskii et al. (2006b).
The quenching function of the magnetic $\alpha$ effect
$\Phi_{_{N}}(B)$ is given by $\Phi_{_{N}}(B) = (3 / 8 B^2) \, [1 -
\arctan (\sqrt{8} B) / \sqrt{8} B] $, where the mean magnetic field
${\bf B}$ is measured in units of the equipartition field $B_{\rm
eq}$ determined by the turbulent kinetic energy, $\Phi_{_{N}}(B) = 1
- (24/5) B^2$ for $B \ll 1/4 $ and $\Phi_{_{N}}(B) = 3/(8B^2)$ for
$B \gg 1/4 $. The nonlinear function $\sigma_{_{B}}(B)$ which is
normalized by $\sigma_{_{B}}(B=0)$, varies from $1$ for $B \ll 1/4$
to $- 11 / 4$ for $B \gg 1/ 4$ (see Rogachevskii and Kleeorin 2004).

Let us consider the simple boundary conditions for a layer of the
thickness $2L$ in the $z$ direction: ${\bf B}(t,|z|=L) = 0$ and
$\chi_c(t,|z|=L) = 0$. We introduce the following non-dimensional
parameters: $D= (l_0 \, S_\ast / L)^2 \, \sigma_{_{B}}(B=0)$ is the
dynamo number and the parameter $S_\ast = S \, L^2 / \eta_{_{T}}$ is
the dimensionless shear number. In the kinematic dynamo, the mean
magnetic field is generated when the dynamo number $D > D_{\rm cr} =
\pi^2/4$ for the symmetric mode (relative to the middle plane $z=0$)
and when the dynamo number $D> D_{\rm cr} = \pi^2$ for the
antisymmetric mode. Numerical solutions of nonlinear
equations~(\ref{RM1})-(\ref{RM3}) have been obtained by Rogachevskii
et al. (2006b). The saturated level of the mean magnetic field
depends strongly on the value of the turbulent diffusivity of the
magnetic helicity $\kappa_{_{T}}$. The mean magnetic field varies
from very small value for $\kappa_{_{T}} = 0.1 \, \eta_{_{T}}$ to
the super-equipartition field for $\kappa_{_{T}} = \eta_{_{T}}$.
This is an indication of very important role of the transport of the
magnetic helicity. The generation of the mean magnetic field causes
negative magnetic $\alpha$ effect, which reduces the growth rate of
large-scale magnetic field. The reason is that the first and the
second terms in the right hand side of Eq.~(\ref{RM1}) have opposite
signs. The first term in Eq.~(\ref{RM1}) describes the shear-current
effect, while the second term in Eq.~(\ref{RM1}) determines the
magnetic $\alpha$ effect. If the magnetic helicity does not
effectively transported out from the generation region, the mean
magnetic field is saturated even at small values of the magnetic
field. Increase of the magnetic helicity flux by increasing of the
turbulent diffusivity $\kappa_{_{T}}$ of magnetic helicity, results
in increase of the saturated level of the mean magnetic field above
the equipartition field. The magnitude of the saturated field
increases also by the increase of the dynamo numbers $D$ within the
range $D_{\rm cr} < D < 2 \, D_{\rm cr}$, and it decreases with the
increase of the dynamo number for $D > 2 \, D_{\rm cr}$. This is a
new feature in the nonlinear mean-field dynamo. For example, in the
nonlinear $\alpha\Omega$ dynamo the saturated level of the mean
magnetic field usually increases with the increase the dynamo
numbers.

The generation of the large-scale magnetic field in a nonhelical
sheared turbulent plasma has been recently investigated by
Brandenburg (2005) using direct numerical simulations (DNS). In
particular, in this DNS the non-convective turbulence is driven by a
forcing that consists of eigenfunctions of the curl operator with
the wavenumbers $4.5 < k_f < 5.5$ and of large-scale component with
wavenumber $k_1=1$. The forcing produces the mean flow $U= U_0 \,
\cos \, (k_1 \, x) \, \cos \, (k_1 \, z)$. The numerical resolution
in these simulations is $128 \times 512 \times 128$ meshpoints, and
the parameters used in these simulations are as following: the
magnetic Reynolds number ${\rm Rm} = u_{\rm rms} / (\eta \, k_f)
=80$, the magnetic Prandtl number ${\rm Pr}_m = \nu/ \eta = 1$ and
$U_0 /u_{\rm rms} =5$. This DNS clearly demonstrate the existence of
the large-scale dynamo in the absence of mean kinetic helicity and
alpha effect. The growth rate of the mean magnetic field is about
$\gamma \, \tau_0 \approx 2 \times 10^{-2}$. This allows us to
estimate the parameter $\sigma_{_{B}}$ characterizing the
shear-current effect, $\sigma_{_{B}} \approx 3.3 \times 10^{-2}$. On
the other hand, our theory predicts $\sigma_{_{B}} = (3 - 6) \times
10^{-2}$ depending on the parameter $\mu$. Note that in DNS by
Brandenburg (2005) the shear is not small (i.e., the parameter $S
\tau_0 \sim 1$), which explains some difference between the
theoretical predictions and numerical simulations. The saturated
level of the mean magnetic field in these numerical simulations is
of the order of the equipartition field which is in a good agreement
with the numerical solutions of the nonlinear dynamo
equations~(\ref{RM1})-(\ref{RM3}) discussed here.

{\it In summary}, we show that in a sheared nonhelical homogeneous
turbulent plasma whereby the kinetic $\alpha$ effect vanishes, the
large-scale magnetic field can grow due to the shear-current effect
from a very small seeding magnetic field.  The dynamo instability is
saturated by the nonlinear effects, and the dynamical nonlinearity
due to the evolution of small-scale magnetic helicity, plays a
crucial role in the nonlinear saturation of the large-scale magnetic
field. Note that a sheared turbulence is a universal feature in
astrophysical plasmas, and the obtained results can be important for
elucidation of origin of the large-scale magnetic fields generated
in astrophysical sheared turbulent plasmas, e.g., in merging
protogalactic clouds or in merging protostellar clouds (Rogachevskii
{\it et al.} 2006a).


\begin{thebibliography}{}

\bibitem[1]{}

    Blackman, E.G. and Field, G., 2002. New dynamical mean-field dynamo
theory and closure approach. Phys. Rev. Lett., 89: 265007 (1-4).

\bibitem[2]{}

    Brandenburg, A., 2005. The case for a distributed solar dynamo
shaped by near-surface shear. Astrophys. J., 625: 539-547.

\bibitem[3]{}

    Brandenburg, A., K\"{a}pyl\"{a}, P. and Mohammed, A., 2004. Non-Fickian
diffusion and tau-approximation from numerical turbulence. Phys.
Fluids, 16: 1020-1027.

\bibitem[4]{}

    Brandenburg, A. and Subramanian, K., 2005a: Astrophysical magnetic fields
and nonlinear dynamo theory. Phys. Rept., 417: 1-209.

\bibitem[5]{}
    Brandenburg, A. and Subramanian, K., 2005b. Minimal tau
approximation and simulations of the alpha effect, Astron.
Astrophys. 439: 835-843.

\bibitem[6]{}

    Brandenburg, A. and Subramanian, K., 2005c. Strong mean field
dynamos require supercritical helicity fluxes. Astron. Nachr., 326:
400-408.

\bibitem[7]{}

    Field, G. and Blackman,  E. G., 2002. Dynamical quenching of the
$\alpha^2$ dynamo. Astrophys. J., 572: 685-692.

\bibitem[8]{}

    Kleeorin, N. and Rogachevskii, I., 1999. Magnetic helicity tensor
for an anisotropic turbulence. Phys. Rev. E, 59: 6724-6729.

\bibitem[9]{}

    R\"{a}dler K.-H., Stepanov R., 2006. Mean electromotive force due to
turbulence of a conducting fluid in the presence of mean flow. Phys.
Rev. E, 73: 056311 (1-15).

\bibitem[10]{}

    Rogachevskii, I. and  Kleeorin, N., 2003. Electromotive force and
large-scale magnetic dynamo in a turbulent flow with a mean shear.
Phys. Rev. E, 68: 036301 (1-12).

\bibitem[11]{}

    Rogachevskii, I. and  Kleeorin, N., 2004. Nonlinear theory of a
"shear-current" effect and mean-field magnetic dynamos. Phys. Rev.
E, 70: 046310 (1-15).


\bibitem[12]{}

    Rogachevskii, I.,  Kleeorin, N., Chernin A. D. and Liverts
E., 2006a. New mechanism of generation of large-scale magnetic
fields in merging protogalactic and protostellar clouds. Astron.
Nachr., 327: 591-594.

\bibitem[13]{}

    Rogachevskii, I.,  Kleeorin, N. and Liverts E., 2006b. Nonlinear
shear-current dynamo and magnetic helicity transport in sheared
turbulence. Geophys. Astrophys. Fluid Dynam., 100: 537-557.

\bibitem[14]{}
Rogachevskii, I. and  Kleeorin, N., 2007. Shear-current effect in a
turbulent convection with a large-scale shear. Phys. Rev. E 75:
046305, (1-7).


\bibitem[15]{}

    R\"{u}diger, G. and Kitchatinov, L.L., 2006. Do mean-field dynamos in
nonrotating turbulent shear-flows exist? Astron. Nachr., 327:
298-303.

\bibitem[16]{}
    Sur, S., Subramanian, K. and Brandenburg, A., 2007. Kinetic and
magnetic alpha effects in nonlinear dynamo theory, Monthly Notices
Roy. Astron. Soc., 376: 1238-1250.


\end{thebibliography}
\end{document}